\documentclass[prb,twocolumn,showpacs,preprintnumbers,amsmath,aps]{revtex4}
\usepackage{graphicx,hyperref}
\usepackage{amssymb}
\usepackage{bm}
\usepackage{subfigure}

\begin{document}

\title{Comment on ``Phase transition for quenched coupled replicas in a plaquette spin model of glasses''}

\author{Giulio Biroli} \email{giulio.biroli@cea.fr}
\affiliation{IPhT, CEA/DSM-CNRS/URA 2306, CEA Saclay, F-91191 Gif-sur-Yvette Cedex, France\\
LPS, Ecole Normale Sup\'erieure, 24 rue Lhomond, 75231 Paris Cedex 05 - France}

\author{Gilles Tarjus} \email{tarjus@lptmc.jussieu.fr}
\affiliation{LPTMC, CNRS-UMR 7600, Universit\'e Pierre et Marie Curie,
bo\^ite 121, 4 Pl. Jussieu, 75252 Paris c\'edex 05, France}

\author{Marco Tarzia} \email{tarzia@lptmc.jussieu.fr}
\affiliation{LPTMC, CNRS-UMR 7600, Universit\'e Pierre et Marie Curie,
bo\^ite 121, 4 Pl. Jussieu, 75252 Paris c\'edex 05, France}

\date{\today}

\maketitle

In a recent letter,\cite{jack_quenched} Jack and Garrahan have studied the thermodynamic phase transition associated with an overlap order parameter in spin models of glasses (PSMs) formed by plaquettes of $p$ spins. The glassy features of such models are known to be be well described in terms of rare defects and kinetic constraints. 
Yet, the authors give strong evidence that the $3$-dimensional square pyramid model (SPyM) 
displays a transition line in the temperature ($T$)-external field ($\epsilon$) plane  which connects a zero-temperature critical point at $T_c(0)=0$ to a finite-temperature one  at $T_c(\epsilon_c)>0$. This result, which complements an earlier study by the same authors\cite{jack_annealed} on the ``annealed'' calculation for two coupled replicas in $d=2$ (triangular plaquette model or TPM) and in $d=3$ (SPyM), shows that the PSMs behave similarly to what expected from the random first-order transition (RFOT) theory, based on the mean-field (MF) theory of the glass transition, with the notable exception that the Kauzmann temperature $T_K=T_c(0)$ is equal to zero.\\ 
Here we show that the properly implemented MF approach for PSMs turns out to be surprisingly predictive for the finite-dimensional behavior. As displayed in Fig. 1, the MF phase diagram of the SPyM on a Bethe lattice with the same connectivity, $c=p=5$, as in $d=3$ is remarkably close to its $d=3$ counterpart in the annealed and the quenched settings.\cite{jack_quenched,jack_annealed}  (It is obtained through standard techniques based on recursion relations and the cavity method.\cite{mezard-parisi01,zdeb10,franz15,BTT}) Both the singular behavior of $T_c(\epsilon)$ and the result $T_K=T_c(0)=0$ are well captured.\cite{footnote-duality} Only, as expected, the MF line terminates at higher $T_c$ and $\epsilon_c$. 
At odds with naive anticipations, the fact that $T_K=0$ is {\it not} the result of long-distance fluctuations but comes from the mere local-connectivity constraint $c=p$: indeed, for $c>p$, MF predicts $T_K>0$.\cite{zdeb10,franz15} 
Note that very poor results (in particular $T_K>0$) would have instead been obtained by using a MF approach based on completely connected lattices. As discussed in Ref. [\onlinecite{CBTT13}], the RFOT found in infinite $d$ is very fragile to short-ranged fluctuations associated with local properties, such as the connectivity of the underlying lattice for spin models. It is therefore crucial to focus on a MF description, such as the Bethe approximation, that takes them into account.\cite{CBTT13} 

All in all, and contrary to expectations, the MF description appears robust with respect to fluctuations as far as overlap properties go.\cite{footnote} Finite-$d$ fluctuations enforce convexity of the potential $V(q)$ (the free-energy cost for maintaining an overlap $q$ with a reference configuration,\cite{FPpotential} sketched in Fig. 1) and prevent true metastability.
Nonetheless, the physics of the unbiased system in $\epsilon=0$ is well described by MF up to a scale, the point-to-set length $\xi_{PTS}$, that diverges as $T \to 0$;\cite{jack_PTS} $V(q)$ remains singular in $d=3$ (with a linear segment) and a configurational entropy $s_c$ can then be univocally defined at low enough $T$, even in finite $d$ (see Fig. 1).
Although one expects $s_c \sim e^{-1/T}$ as $T\to T_K=0$ in both MF and $d=3$,\cite{jack_quenched} other critical exponents are likely to be different, as, {\it e.g.}, that characterizing the relation between $\xi_{PTS}$ and $s_c$. Given the robustness of the thermodynamic MF predictions, understanding the implications for the dynamics and the connection to the defect picture is a key question.\cite{foini12} Further studies of the PSMs could be instrumental in clarifying it.  
\begin{figure}[h]
\includegraphics[width=\linewidth]{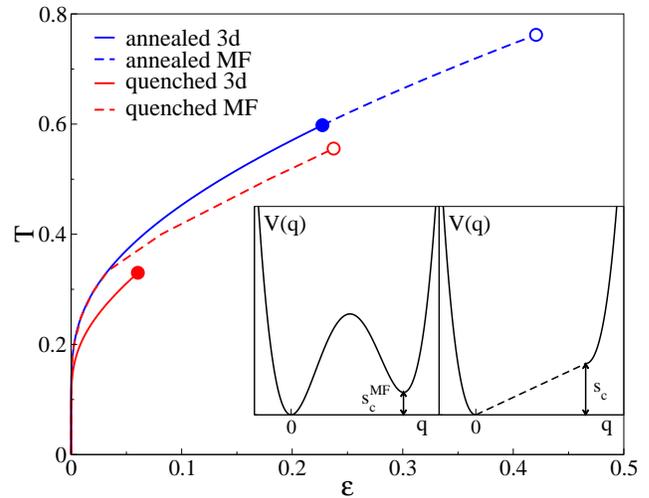}
\caption{Phase diagram of the SPyM in both the annealed and the quenched descriptions for the Bethe lattice with $c=5$ (dashed lines) and for $d=3$ (full lines\cite{jack_quenched,jack_annealed}). Inset: Sketch of the potential $V(q)$ in MF (left) and for $d=3$ (right).}
\label{fig1}
\end{figure}

\end{document}